# Effects of nongaussian diffusion on "isotropic diffusion" measurements: an ex-vivo microimaging and simulation study

Sune Nørhøj Jespersen[1,2,*], Jonas Lynge Olesen[1,2], Andrada Ianuş[3,4], Noam Shemesh[3]

Author affiliations

[1]Center of Functionally Integrative Neuroscience (CFIN) and MINDLab, Department of Clinical Medicine, Aarhus University, Aarhus, Denmark.

[2]Department of Physics and Astronomy, Aarhus University, Aarhus, Denmark.

[3]Champalimaud Neuroscience Programme, Lisbon, Portugal

[4]Center for Medical Image Computing, Department of Computer Science, University College London, London, United Kingdom

[*]Corresponding author:

Sune Nørhøj Jespersen

CFIN/MindLab and Dept. of Physics and Astronomy, Aarhus University

Nørrebrogade 44, bygn 10G, 5. sal

8000 Århus C

Denmark

Cell: +45 60896642

E-mail: sune@cfin.au.dk

**Abstract**

Designing novel diffusion-weighted pulse sequences to probe tissue microstructure beyond the conventional Stejskal-Tanner family is currently of broad interest. One such technique, multidimensional diffusion MRI, has been recently proposed to afford model-free decomposition of diffusion signal kurtosis into terms originating from either ensemble variance of isotropic diffusivity or microscopic diffusion anisotropy. This ability rests on the assumption that diffusion can be described as a sum of multiple Gaussian compartments, but this is often not strictly fulfilled. The effects of nongaussian diffusion on single shot isotropic diffusion sequences were first considered in detail by de Swiet and Mitra in 1996. They showed theoretically that anisotropic compartments lead to anisotropic time dependence of the diffusion tensors, which causes the measured isotropic diffusivity to depend on gradient frame orientation. Here we show how such deviations from the multiple Gaussian compartments assumption conflates orientation dispersion with ensemble variance in isotropic diffusivity. Second, we consider additional contributions to the apparent variance in isotropic diffusivity arising due to intracompartmental kurtosis. These will likewise depend on gradient frame orientation. We illustrate the potential importance of these confounds with analytical expressions, numerical simulations in simple model geometries, and microimaging experiments in fixed spinal cord using isotropic diffusion encoding waveforms with 7.5 ms duration and 3000 mT/m maximum amplitude.

# Introduction

The classical Stejskal-Tanner diffusion NMR method [1] has been instrumental throughout disciplines from porous media to biomedical imaging. Motivated by the need to obtain information which is independent or even unattainable from conventional diffusion weighting [2, 3], numerous approaches probing the behavior of diffusing spins with "orthogonal measurements" have been proposed [3-9]. Such information has proved necessary to inform (bio)physical modeling of tissue microstructure, and to constrain parameter estimation in those models [10, 11]. Additionally, under some circumstances, certain sample characteristics can be quantified in a model free fashion using generalized diffusion encoding sequences. For example, so-called multiple diffusion encoding sequences, have been shown to afford quantification of microscopic diffusion anisotropy and diffusion tensor variance based only on an assumption of multiple non-exchanging compartments [4, 12-17]. Other designs of so-called multidimensional diffusion pulse sequences have been used to quantify various aspects of the diffusion tensor distribution in a system of multiple Gaussian compartments [18]. The assumption of multiple Gaussian compartments is also a commonplace assumption in diffusion modeling, see e.g. [9, 19] and references therein. In such systems, net kurtosis arises exclusively due to ensemble variance in diffusion tensor properties [20], and the analysis elegantly exploits analogies between diffusion in Gaussian compartments and chemical shift anisotropy in solid-state NMR [5] to decompose this variance into distinct sources purportedly related to microstructure. For example, by varying the shape of the diffusion gradient waveform, it becomes possible to disentangle microscopic diffusion anisotropy from variance in isotropic diffusivity [21].

However, in many porous and biologically relevant systems, where such methods could be of great interest, approximating the signal as arising from multiple Gaussian compartments is strictly speaking justified only in very specific regimes, typically when small diffusion weighting is used, or short or long diffusion times are applied. Indeed, multiple reports have demonstrated pronounced time or frequency dependencies of the diffusion coefficient in biological tissue [22-34]. Hence, the key assumption of current multidimensional diffusion MR is likely to be violated, at least in principle, during part of their inherently long [35] diffusion gradient application. This has at least two consequences, which complicates the interpretation of data derived from such classes of sequences: (1) anisotropic time dependence of diffusion and kurtosis leads to orientation, and therefore dispersion, dependence, and (2) the intracompartmental kurtosis, i.e. kurtosis for a single compartment, is non-vanishing. The former effect, the orientation dependence of isotropic diffusivity due to anisotropic time dependence, was first identified and discussed in depth by de Swiet and Mitra [36]. Here, we study the implications of nongaussian diffusion for multidimensional diffusion weighting and show in particular that both phenomena have consequences for accurate identification of kurtosis with the variance of diffusion tensor properties over the ensemble. We present analytical calculations supplemented with numerical simulations to examine the magnitude of deleterious effects in various geometries, and support our findings with diffusion experiments in fixed pig and rat spinal cords.

## Theory

Interpretation of multidimensional diffusion MRI is based on the multiple Gaussian compartments assumption. We first give a simplified account of the principles of multidimensional diffusion encoding under this assumption. For more detail, see the recent review by Topgaard [18]. We then derive the consequences of violating the Gaussian assumption within compartments. We use the terms compartments, pores, and microdomains interchangeably for spatial regions with negligible exchange.

*Multiple Gaussian Compartments*

For diffusion in a Gaussian compartment with diffusion tensor $\mathbf{D}$, the diffusion signal $S(\mathbf{b})$ normalized so $S(\mathbf{b}=0)=1$, reads [37, 38]

$$S(\mathbf{b}) = e^{-b_{ij}D_{ij}}, \qquad (1)$$

where we use the convention of sum over repeated indices, and the matrix $\mathbf{b}$ encapsulates the diffusion weighting

$$b_{ij} = \int_0^T dt\, q_i(t)q_j(t) = \int_{-\infty}^{\infty} \frac{d\omega}{2\pi} q_i(\omega)q_j^*(\omega) \qquad (2)$$

played out by the diffusion gradients $\mathbf{G}(t)$: In Eq. (2), T is the measurement time and $\mathbf{q}(\omega)$ is the Fourier transform of the diffusion wave vector (gradient area) defined by

$$\mathbf{q}(t) \equiv \gamma \int_0^t ds\, \mathbf{G}(s)$$

For a collection of non-communicating compartments (e.g. pores or cells), the net diffusion signal is a sum over the $N$ individual pore signals, each having their own (microscopic) diffusion tensor $\mathbf{D}^{(p)}$ (note that $p$ is not a power, but a label for pore $p$):

$$S(\mathbf{b}) = \frac{1}{N}\sum_p e^{-b_{ij}D_{ij}^{(p)}} = \left\langle \exp(-b_{ij}D_{ij}^{(p)}) \right\rangle \qquad (3)$$

Here we introduce the shorthand $\langle \cdot \rangle$ for the average over compartments. Because of this sum, the net signal becomes nongaussian (unless the microscopic diffusion tensors happen to be all identical), yet time independent, and can be expanded using the cumulant expansion for small $b$ to yield [39-42]

$$\log S(b) \approx -b_{ij}\langle D_{ij}^{(p)}\rangle + \frac{1}{2}b_{ij}b_{kl}\left(\langle D_{ij}^{(p)}D_{kl}^{(p)}\rangle - \langle D_{ij}^{(p)}\rangle\langle D_{kl}^{(p)}\rangle\right) \quad (4)$$

While the first term for arbitrary b-matrix can be reproduced by combining several single diffusion encoding (e.g. Stejskal-Tanner) measurements [2, 43], the second term ("kurtosis term") is of greater interest here, because it explicitly yields the variance of diffusion tensor properties over the ensemble.

The idea of the multidimensional diffusion encodings is that $\mathbf{G}(t)$ can be designed to produce a b-matrix via Eq. (2) that probes various aspects of the Gaussian compartment. For example, for isotropic diffusion encoding [5, 21, 44, 45], $\mathbf{G}(t)$ is chosen to give $b_{ij} = \delta_{ij}b/3$, with

$$b \equiv \mathrm{Tr}(\mathbf{b}) = \sum_i \int_0^T dt\, q_i^2(t),$$

which leads immediately to $b_{ij}D_{ij}^{(p)} = bD_I^{(p)}$, where $D_I^{(p)} = \mathrm{Tr}(\mathbf{D}^{(p)})/3$ denotes the mean or isotropic diffusivity in pore $p$. Hence the measurement returns

$$\log S_I(\mathbf{b}) \approx -b\langle D_I^{(p)}\rangle + \frac{1}{2}b^2\left(\langle (D_I^{(p)})^2\rangle - \langle D_I^{(p)}\rangle^2\right) \equiv -bD_I + \frac{1}{2}b^2 V_I \quad (5)$$

i.e., the ensemble average of isotropic diffusivity $D_I$ ($b$-term) and its variance $V_I = \mathrm{Var}(D_I^{(p)})$ ($b^2$-term). Following [46], we can use the diffusion kurtosis imaging (DKI) formalism [41]

$$\log S_I \approx -bD_I + \frac{1}{6}b^2 D_I^2 K_I, \quad (6)$$

to relate the kurtosis $K_I$ from the isotropic diffusion weighting to the variance in isotropic diffusivity $V_I$

$$K_I = 3V_I / D_I^2 \quad (7)$$

It is more convenient to report the dimensionless $K_I$ than $V_I$ due to the normalization. Typical values of $K_I$ in the human brain reported in [8, 46, 47] fall in the range of 0.25-0.6, although signal decay appears close to monoexponential [47, 48].

These ideas can be extended to other waveforms designed to probe the variance of other aspects of $\mathbf{D}$, such as e.g. its asymmetry and anisotropy [18]: In theory, the entire distribution $P(\mathrm{D})$ can be reconstructed with sufficient sampling of $S$. However, as we show below, these useful properties break down for nongaussian diffusion.

*Non-Gaussian diffusion effects*

A general signal expression (ignoring relaxation) for diffusion within a single pore is given by [42, 49]

$$S = \left\langle \exp\left( i\int_0^T \mathbf{q}(t) \cdot \mathbf{v}(t) dt \right) \right\rangle \equiv \left\langle \exp(i\phi) \right\rangle \tag{8}$$

where $\langle \cdot \rangle$ includes an average over the trajectories $\mathbf{r}(t)$ of all contributing spins and $\phi = \int_0^T \mathbf{q}(t) \cdot \mathbf{v}(t) dt$ is the phase accrued for a spin with velocity $\mathbf{v} = \dot{\mathbf{r}}(t)$. Considering that $b$ is essentially proportional to the amplitude of the diffusion gradient $\mathbf{G}$ squared, we can expand Eq. (8) to second order in $b$, giving

$$S \approx \exp\left( -\frac{1}{2}\langle \phi^2 \rangle + \frac{1}{4!}\left( \langle \phi^4 \rangle - 3\langle \phi^2 \rangle^2 \right) \right). \tag{9}$$

From this we make 2 observations of relevance to multidimensional diffusion encoding:

a) The first order term

$$\frac{1}{2}\langle \phi^2 \rangle = \int \frac{d\omega}{2\pi} q_i(\omega) q_j^*(\omega) \mathcal{D}_{ij}(\omega) \tag{10}$$

with $\mathcal{D}_{ij}(\omega)$ the fourier transform of $\mathcal{D}_{ij}(t) \equiv \theta(t)\langle v_i(t) v_j(0) \rangle$ [49] is explicitly sensitive to the time dependence of the diffusion tensor D. For true Gaussian diffusion, diffusion is frequency or time independent, and $\mathcal{D}_{ij}(\omega) = D_0$ can be moved outside the integral, as assumed in the analysis of multidimensional diffusion encoding. In general, however, D is expected to be time or frequency dependent in most practical experimental ranges. Importantly, each component $D_{ij}$ depends on time or frequency in a different way for anisotropic compartments. It follows that $D_I$ from each anisotropic compartment depends on its orientation relative to the gradient frame [36], thereby conflating the ensemble variance of isotropic diffusivity $D_I$ with dispersion. Extending this line of thought, the 4$^{th}$ order term, which involves the time-dependent kurtosis tensor, will also depend on gradient frame orientation for anisotropic domains.

b) Even the single compartment signal expression in Eq. (9) acquires contributions at the level of $b^2$, which add to the population variance contribution derived above in Eq. (5), and hence in principle prevents a clean interpretation in terms of e.g. isotropic variance.

For concreteness, we focus in what follows on magic angle spinning of the q-vector (q-MAS) for isotropic diffusion encoding and implement the piecewise harmonic diffusion gradient waveform described in [50], with total duration $T = 2\tau$. However, the principles apply equally to other generalized diffusion gradient waveforms.

*Time-dependent diffusivity*

For simplicity, we make the assumption that the diffusion tensor eigenvectors are time independent, such that $\mathcal{D}$ can be expanded in terms of its 3 eigenvectors $\hat{\mathbf{v}}^k$ ($k = 1, 2, 3$) and associated (frequency dependent) eigenvalues $\lambda_k(\omega)$

$$\mathcal{D}_{ij}(\omega) = \hat{\mathbf{v}}_i^k \hat{\mathbf{v}}_j^k \lambda_k(\omega) \tag{11}$$

The *apparent* isotropic diffusivity $\tilde{D}_I$ is found from

$$b\tilde{D}_I = \frac{1}{2}\langle \phi^2 \rangle$$

When inserting Eq. (11) into Eq. (10), we then find

$$b\tilde{D}_I = \hat{\mathbf{v}}_i^k \hat{\mathbf{v}}_j^k \int \frac{d\omega}{2\pi} q_i(\omega) q_j^*(\omega) \lambda_k(\omega). \tag{12}$$

Because each of the three eigenvalues in general have different frequency dependencies (e.g. anisotropy), the three terms multiplying the eigenvectors in Eq. (12) are different, and therefore $\tilde{D}_I$ depends on the orientation of the diffusion tensor eigenframe relative to the laboratory frame defined by the diffusion gradients. Hence as noted in [36], the derived property will not be rotationally invariant, an unfortunate theoretical drawback for the quantification of e.g. isotropic diffusivity.

As a more concrete example, we specialize to the case of a single pore with axial symmetry andsymmetry axis by $\hat{\mathbf{u}}$. In this case, we find from Eq. (12)

$$b\tilde{D}_I = \int \frac{d\omega}{2\pi} |\mathbf{q}(\omega)|^2 \mathcal{D}_\perp(\omega) - \hat{u}_i \hat{u}_j \int \frac{d\omega}{2\pi} q_i(\omega) q_j^*(\omega) (\mathcal{D}_\parallel(\omega) - \mathcal{D}_\perp(\omega)). \tag{13}$$

Introducing the notation

$$\mathcal{B}_{ij}^{\perp} = \int \frac{d\omega}{2\pi} q_i(\omega) q_j^*(\omega) \mathcal{D}_{\perp}(\omega) \tag{14}$$

$$\mathcal{B}_{ij}^{\Delta} = \int \frac{d\omega}{2\pi} q_i(\omega) q_j^*(\omega) (\mathcal{D}_{\parallel}(\omega) - \mathcal{D}_{\perp}(\omega)), \tag{15}$$

Eq. (13) becomes the quadratic form

$$\tilde{D}_I = \delta_{ij} \mathcal{B}_{ij}^{\perp} + u_i u_j \mathcal{B}_{ij}^{\Delta} = \text{Tr}(\mathcal{B}^{\perp}) + \hat{\mathbf{u}}^T \mathcal{B}^{\Delta} \hat{\mathbf{u}}. \tag{16}$$

with an explicit dependence on the orientation $\hat{\mathbf{u}}$ of the compartment. Similar results were obtained first by de Swiet and Mitra using a different notation [36]. We next consider the consequences for the kurtosis.

We note that the $\tilde{D}_I$ takes on its extreme values when $\hat{\mathbf{u}}$ is aligned with the principal eigenvector of $\mathcal{B}^{\Delta}$ (maximum) and with the eigenvector associated with the smallest eigenvalue of $\mathcal{B}^{\Delta}$ (minimum). Thus, the orientational variability of our measurement is dictated by the range of the eigenvalues of $\mathcal{B}^{\Delta}$. In general, the sample will feature some distribution $\mathcal{P}(\hat{\mathbf{u}})$ of domain orientations $\hat{\mathbf{u}}$ causing the measured isotropic diffusivity $\tilde{D}_I$ to reflect not only isotropic diffusivities, but also orientation dispersion, counteracting the goal of separating orientation dispersion from heterogeneity in isotropic diffusivity. We label the additional contribution to the isotropic kurtosis originating from dispersion as $\delta K_I^{(d)}$. Specifically, if the sample contains pores differing only in the orientation, the entire variance in apparent isotropic diffusivities

$$\frac{1}{3} \tilde{D}_I^2 \delta K_I^{(d)} = \left\langle (\tilde{D}_I^{(p)})^2 \right\rangle - \left\langle \tilde{D}_I^{(p)} \right\rangle^2 = \mathcal{B}_{ij}^{\Delta} \mathcal{B}_{lk}^{\Delta} \left( \left\langle u_i u_j u_k u_l \right\rangle - \left\langle u_i u_j \right\rangle \left\langle u_k u_l \right\rangle \right) \tag{17}$$

is due to orientation dispersion $\left\langle u_i u_j u_k u_l \right\rangle - \left\langle u_i u_j \right\rangle \left\langle u_k u_l \right\rangle$. Specifically, in the case of a powder average of identical pores, $\mathcal{P}(\hat{\mathbf{u}}) = 1/4\pi$, the variance in apparent isotropic diffusivity becomes

$$\frac{1}{3} \tilde{D}_I^2 \delta K_I^{(d)} = \frac{2}{5} \text{Var}(b_i^{\Delta}), \tag{18}$$

where $b_i^{\Delta}$ are the eigenvalues of $\mathcal{B}^{\Delta}$. The true ensemble variance of the isotropic diffusivity is 0.

*Fourth order cumulant*

Equation (6) for the q-MAS experiment acquired in a single compartment reads

$$\log S \approx -b \tilde{D}_I + \frac{1}{6} b^2 \tilde{D}_I^2 \tilde{K}_I, \tag{19}$$

where $\tilde{K}_I$ is the apparent kurtosis, depending on $\mathbf{G}(t)$ just as $\tilde{D}_I$. With a population of pores with each of their apparent isotropic diffusivities $\tilde{D}_I^{(p)}$ and kurtoses $\tilde{K}_I^{(p)}$, the net signal becomes

$$S(b) = \sum_p e^{-b\tilde{D}_I^{(p)} + \frac{1}{6}b^2(\tilde{D}_I^{(p)})^2 \tilde{K}_I^{(p)} + \mathcal{O}(b^3)} \quad (20)$$

keeping only up to second-order in $b$ in each pore. For the net signal, this then amounts to

$$\log S_I(b) = -b\langle \tilde{D}_I^{(p)} \rangle + \frac{1}{2}b^2 \left( \frac{1}{3}\langle (\tilde{D}_I^{(p)})^2 \tilde{K}_I^{(p)} \rangle + \langle (\tilde{D}_I^{(p)})^2 \rangle - \langle \tilde{D}_I^{(p)} \rangle^2 \right) + \mathcal{O}(b^3) \quad (21)$$

The second-order term no longer just reflects the variance in $\tilde{D}_I^{(p)}$, but acquires an additional contribution $\delta K_I^{(i)} = \langle (\tilde{D}_I^{(p)})^2 \tilde{K}_I^{(p)} \rangle / \langle \tilde{D}_I^{(p)} \rangle^2$ from the average intracompartmental kurtosis [47]. The relative importance of this term can obviously not be made small by decreasing b. Specifically, we can deduce a non-vanishing ensemble variance in isotropic diffusivity in the presence of only a single compartment.

## Simulations

For our numerical examples, we model the time dependence of the diffusivity based on diffusion within a one-dimensional interval of length $a$ and with bulk diffusivity $D_0$. In this case, the propagator is known exactly, and the diffusivity $D(t) \equiv D_0 d(t/t_D)$ as function of time ($t_D \equiv a^2/D_0$) can be computed

$$d(s) = \frac{1}{12s} - \frac{8}{\pi^4 s} \sum_{n=0}^{\infty} \frac{1}{(2n+1)^4} \exp\left(-s\pi^2(2n+1)^2\right).$$

Building on this expression, we model two types of compartment geometries, both based on a rectangular box with side lengths $a_\parallel$ and $a_\perp$: we loosely refer to them as a "square cylinder" for $a_\parallel = \infty$ and "box" for finite $a_\parallel$ and $a_\perp$. For the square cylinder

$$D_\parallel(t) = D_0, \quad D_\perp = D_0 d(t/t_D) \quad (22)$$

while for the box

$$D_\parallel(t) = D_0 d(D_0 t / a_\parallel^2), \quad D_\perp = D_0 d(D_0 t / a_\perp^2) \quad (23)$$

When examining the orientational variability of $\tilde{D}_I$, we plot the maximum $\max(\tilde{D}_I)$ and minimum $\min(\tilde{D}_I)$ values of $\tilde{D}_I$ corresponding to the major and minor eigenvectors of $\mathcal{B}^\Delta$, as well as the normalized variance of these 2 numbers, i.e. $(\max(\tilde{D}_I) - \min(\tilde{D}_I))^2 / (\max(\tilde{D}_I) + \min(\tilde{D}_I))^2$, as function of pore size. Since this is precisely the variance $(\max(\tilde{D}_I) - \min(\tilde{D}_I))^2 / 4$ divided by the squared mean $(\max(\tilde{D}_I) + \min(\tilde{D}_I))^2 / 4$ of isotropic diffusivities in a system consisting of two pores pointing along the eigenvector directions of the maximum and minimum eigenvalues of $\mathcal{B}^\Delta$, it is equal to $(1/3)\tilde{D}_I^2 \delta K_I^{(d)}$ for such a system, as per the definition in Eq. (17). We also compare with the "true isotropic diffusivity"

$$D_I(t_{eff}) = \frac{1}{3} D_\parallel(t_{eff}) + \frac{2}{3} D_\perp(t_{eff}) \tag{24}$$

at the effective diffusion time $t_{eff}$ given by $t_{eff} \equiv b/q^2$ [5]. Note however that the diffusion time is not uniquely defined for general gradient waveforms. In addition, we plot $\delta K_I^{(d)}$ for a powder average, Eq. (18).

For the intracompartmental kurtosis contribution, we have the apparent isotropic kurtosis in terms of the phase moments

$$\tilde{D}_I^2 \tilde{K}_I = \frac{1}{4}\left(\langle \phi^4 \rangle - 3\langle \phi^2 \rangle^2 \right)/b^2 = \frac{\langle \phi^4 \rangle}{4b^2} - 3\tilde{D}_I^2 \tag{25}$$

Although an analytical expression for $\langle \phi^4 \rangle$ can be found in the stick, we here compute it numerically using Monte Carlo simulations with 7 million particles. The 4$^{th}$ order term is more difficult to estimate robustly with numerical simulations, and therefore we here consider two slightly simpler systems, a one-dimensional finite stick (length $a$) and a one-dimensional line of regularly spaced (lattice spacing $a$) permeable membranes, having permeability $\kappa = 0.02$ µm/ms. The latter system was included to have a system in which the diffusivity, and thereby $D^2K$ does not go to zero asymptotically, as for restricted geometries. We create two ensembles for each geometry, i) all lines parallel to $\hat{z}$, for which

$$\langle \phi^4 \rangle = \int_0^T dt_1 \ldots \int_0^T dt_4 G_z(t_1) G_z(t_2) G_z(t_3) G_z(t_4) \langle z(t_1) z(t_2) z(t_3) z(t_4) \rangle \tag{26}$$

and ii) the corresponding powder for which symmetry considerations yield

$$\langle \phi^4 \rangle = \frac{1}{5} \int_0^T dt_1 \ldots \int_0^T dt_4 \mathbf{G}(t_1) \cdot \mathbf{G}(t_2) \mathbf{G}(t_3) \cdot \mathbf{G}(t_4) \langle z(t_1) z(t_2) z(t_3) z(t_4) \rangle \tag{27}$$

For both systems, obviously $\langle (D_I^{(p)})^2 \rangle - \langle D_I^{(p)} \rangle^2 = 0$, since the pores are identical and only differ in orientation, and true isotropic diffusivity is orientationally invariant. We examine the magnitude of the inferred variance $\delta K_I^{(i)}$ as function of $\tau/t_D$.

In addition to these geometric model cases, we also use the experimental time-dependent diffusivities found in fixed pig spinal cord white matter reported recently [23]. There, modeling was used to infer the time-dependent compartment diffusivities $D_a$ for intra-axonal axial diffusion, and $D_{e,\|}$ and $D_{e,\perp}$ for extra-axonal axial and radial diffusion. Using Eqs. (14)-(16), the outcome of the q-MAS experiment is estimated by putting $D_\|(t) = D_a(t)$ and $D_\perp = 0$ for the intra-axonal compartment, and $D_\|(t) = D_{e,\|}(t)$ and $D_\perp = D_{e,\perp}(t)$ for the extra-axonal compartment. The "true isotropic diffusivity" corresponding to a particular waveform duration $\tau$ is estimated by interpolating the time dependent compartmental diffusivities to the effective diffusion time $t_{eff}$ [5]

$$D_I(t_{eff}) = \frac{1}{3}\left( f D_a(t_{eff}) + (1-f) D_{e,\|}(t_{eff}) \right) + \frac{2}{3} D_{e,\perp}(t_{eff}). \tag{28}$$

## Experiments

All animal experiments were preapproved by the local ethics committee and fully complied with local and EU laws. Cervical spinal cord specimens were obtained from n = 2 rats using standard perfusion procedures. Briefly, rats were deeply anesthetized with pentobarbital and then the rats were perfused transcardially. After fixation, the tissues were immersed in 4% PFA for 24h post-mortem, and then placed in a PBS solution for additional 24h. The specimens were mounted in a 5 mm NMR tube filled with Fluorinert (Sigma Aldrich, Lisbon, Portugal). Imaging was performed on a 16.4T Bruker Aeon Ascend magnet interfaced with an AVANCE IIIHD console, and equipped with a micro5 probe with gradients capable of producing up to 3000 mT/m in all directions. A birdcage coil with inner diameter of 5 mm, tuned for 1H was used for both transmission and signal reception. A spin-echo EPI sequence was modified to accommodate isotropic diffusion encoding gradient waveforms, which were generated according to [50, 51] with a duration of 7.5 ms and separation of 1.15 ms. The following acquisition parameters were used: EPI bandwidth = 441kHz, 2-shots, matrix size 80 by 60 with, in-plane resolution = 75 by 75 , slice thickness = 1.8 mm, number of averages = 4, and TR/TE = 3750 / 41 ms. The isotropic diffusion gradient sequence was acquired with 20 b-values ranging linearly from 0 to 3 ms/$\mu$m$^2$, and 12 different orientations were acquired for each b-value. This scheme was then repeated 4 times to allow comparison of variability over orientations with measurement uncertainty. A measure of the average SNR was computed as the mean signal in the entire spinal cord divided by the voxel-wise standard deviation over the 960 $b=0$ images. SNR calculated this way was approximately 101 for b = 0, and 31 for b = 3 ms/$\mu$m$^2$. The apparent isotropic diffusivity was then estimated by fitting to the DKI expression for each orientation separately. The parameters were initialized by the estimates of a prior DKI fit to independent data acquired with a standard EPI DTI sequence (6 b = 0 images, 21 b-values from 0 to 3 ms/, 30 directions).

To rule out experimental artifacts, an additional experiment was performed in a solution of PVP40 (Polyvinylpyrrolidone, Sigma Aldrich, Lisbon, Portugal) with a mass concentration of 40% in a mixture of H2O and D2O (1:9), which has similar diffusivity to ex-vivo tissue.

All data from this study is available for free download from https://github.com/sunenj/.

## Results

*Simulations*

In Figure 1a, we demonstrate the variability of $\tilde{D}_I$ in the "square cylinder" depending on its orientation. The black lines refer to the left-hand y-axis and plot the minimum and maximum values as a function of $a$ when $\tau = 55$ ms as in [52]. The upper x-axis gives the corresponding relative timescale $\tau/t_D$, the dimensionless parameter which fully characterizes the diffusion time and pore size dependence. This allows to readily translate to other waveform durations, by calculating the ratio $\tau/t_D$ of interest. The graph also plots the true isotropic diffusivity for context. For very small compartments or large $\tau$, only the axial contribution matters, and the apparent isotropic diffusivity approaches the true isotropic diffusivity of $D_0/3$, regardless of orientation. The orientational dependence then increases with compartments size, reaching about 25% of average $\tilde{D}_I$. To convey the importance for the kurtosis contribution of this variability in $\tilde{D}_I$ due to orientation, the blue lines plot the corresponding normalized variance. It reaches 0.04 when the waveform duration is close to the diffusion length, but remains below 0.01 for $a \lesssim 5$ μm. Figure 1b demonstrates $\delta K_I^{(d)}$ for the corresponding powder, revealing an isotropic kurtosis contribution due to dispersion of up to 0.012.

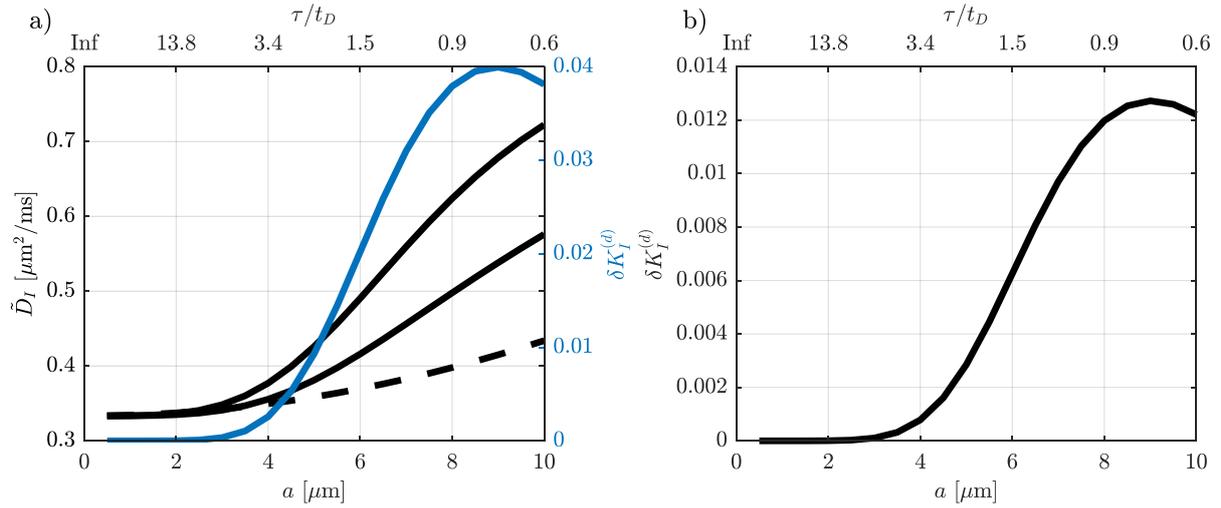

**Figure 1** *Results for $\tilde{D}_I$ in the square cylinder. In (a), black solid lines plot the maximum and mininum $\tilde{D}_I$ over directions as function of cylinder radius (bottom) or $\tau/t_D$ (top). The dashed curve is the true isotropic diffusivity at an effective diffusion time $t_{eff} = b/q^2$. The blue curve refers to the right hand y-axis, and gives the corresponding normalized variance. In (b), the standard deviation of $\tilde{D}_I$ over a powder distribution of identical square cylinders is shown.*

Figure 2 examines the orientational variability of $\tilde{D}_I$ for the box system. In a), we show $\delta K_I^{(d)}$ function of box side lengths $a_\parallel$ (left y-axis) and $a_\perp$ (lower x-axis) when $\tau = 55$ ms, with corresponding values of $\tau/t_{D_\parallel}$ (right y-axis) and $\tau/t_{D_\perp}$ (upper x-axis). The dashed red line (full red line) marks the region in which $\delta K_I^{(d)}$ exceeds 0.1 (0.05), corresponding to 17–40% (8–20%) of reported $K_I$ values [8, 46, 47]. We see that this is a rather extended region for this system, limiting the applicability to a region corresponding to highly isotropic domains, particularly for the smaller box dimensions. Figure 2b maps $\delta K_I^{(d)}$ for the corresponding powder system, showing a somewhat larger tolerance for anisotropy.

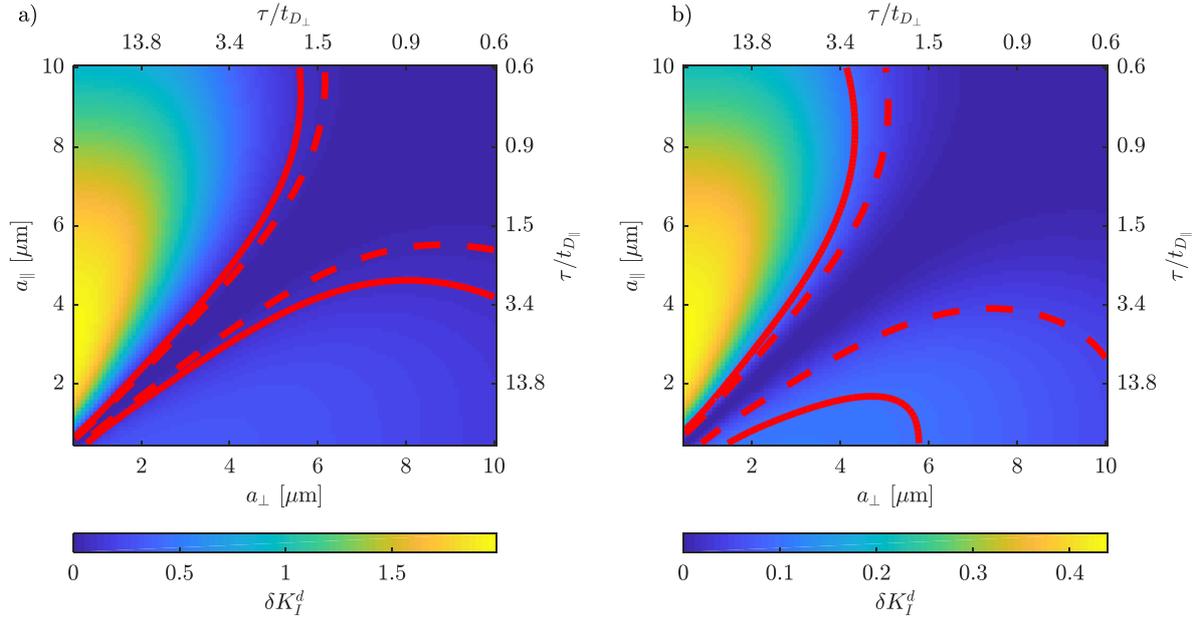

**Figure 2** *Color coded results for $\delta K_I^{(d)}$ in a rectangular box as function of side lengths or $\tau/t_D$. In (a), $\delta K_I^{(d)}$ for a system consisting of two rectangular boxes is shown, whereas the powder system is shown in (b). Red (dashed) lines delineate regions where the difference is below 0.1 (0.05).*

In Figure 3, we have inferred time-dependent diffusivities from [23] in intra- and extra-axonal spaces to compute the isotropic diffusivity as function of waveform duration which would be observed with q-MAS experiment of varying duration according to Eq. (13). Clearly, there is a pronounced orientational dependence, especially for the intra-axonal space, which is also the most anisotropic space. Here, the dispersion contribution to isotropic kurtosis plateaus around 0.015 for waveforms longer than 40 ms. For comparison, the dashed lines show the actual isotropic diffusivity computed from the compartmental diffusivities interpolated to the effective diffusion time $t_{\text{diff}} = b/q^2$. For both the intra-and extra-axonal spaces, the apparent isotropic diffusivities consistently overestimate the true isotropic diffusivity, regardless of gradient orientation. Figure 3c is similar to a and b, but combines the intra- and extra-axonal spaces according to the volume fraction $f \approx 0.5$ found in [23].

This attenuates the relative range somewhat compared to the intra-axonal space, although $\tilde{D}_I$ still can vary more than 10% of its mean and is consistently overestimated compared to the true isotropic diffusivity. However, the reduced orientational variability of apparent isotropic diffusivity for the two compartment system compared to each of the compartments individually does not necessarily imply that the contamination of isotropic variance by dispersion is small. For a system with two anisotropic Gaussian compartments with identical orientation distribution function $\mathcal{P}(\hat{\mathbf{u}})$, one can show that the net kurtosis has three terms

$$\tilde{K}_I \tilde{D}_I^2 = \left( (1-f)\delta K_{I,e}^{(d)} \tilde{D}_{I,e}^2 + f \delta K_{I,a}^{(d)} \tilde{D}_{I,a}^2 + 3f(1-f)\left(\langle \tilde{D}_{I,e} \rangle - \langle \tilde{D}_{I,a} \rangle\right)^2 \right) \qquad (29)$$

where the first two terms originate from the dispersion variance Eq. (17) in each of the compartments (extra and intra, respectively), and the last term comes from the ensemble variance of mean apparent isotropic diffusivity. The blue curve in Figure 3c shows the kurtosis contributed by orientation dispersion relative to the total apparent isotropic kurtosis Eq. (29) for a powder sample consisting of the intra-and extra axonal compartments. For most of the time range in this system, it exceeds 20%. This large fraction is partly due to the near equality of the compartment mean diffusivities, as has also been found by other authors [47, 53-55], and the situation may be more favorable in other systems.

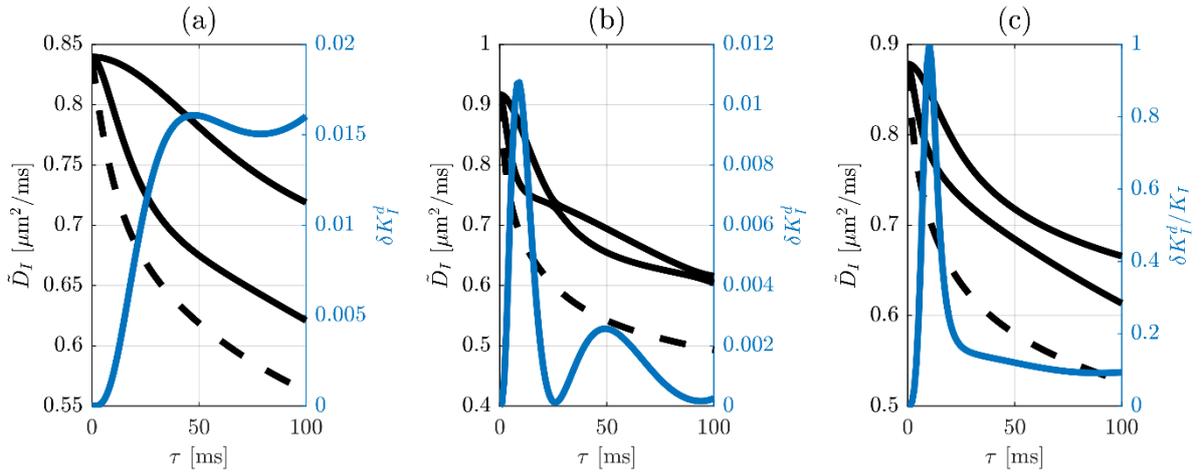

**Figure 3** *Maximum and mininum $\tilde{D}_I$ (black) and their corresponding $\delta K_I^{(d)}$ (blue) for intra (a) and extra (b) axonal compartments in fixed pig spinal cord as function of $\tau$. The dashed lines trace out $D_I(t_{\text{diff}})$ at the effective diffusion time $t_{\text{diff}} = b/q^2$ [5]. Figure 3c shows maximum and minimum $\tilde{D}_I$ for the combined intra-and extra-axonal system (black solid line) compared to true $D_I$ (black dashed line). The blue curve plots the apparent isotropic kurtosis originating from dispersion in a powder distribution relative to the total apparent isotropic kurtosis. Data from Fasiculus Cuneatis (ROI C) in [23].*

In Figure 4, we examine the orientational variability of the apparent isotropic kurtosis (still ignoring intracompartmental kurtosis) for a two compartment system of parallel sticks (axons, no dispersion) and a cylindrical extra axonal diffusion tensor, i.e.

$$\tilde{D}_I^2 \tilde{K}_I = 3f(1-f)(\tilde{D}_{I,e} - \tilde{D}_{I,a})^2 \qquad (30)$$

as a function of $\tau$. The compartment diffusivities are those obtained from the 7 white matter ROIs in [23]. The gray shaded area in the figure outlines the possible outcomes of Eq. (30) as the orientation of the gradient frame is varied, and the dashed orange line is the average of those values. This is then compared to the ground truth

$$D_I^2 K_I = 3f(1-f)(D_{I,e} - D_{I,a})^2 \qquad (31)$$

(green line) at the corresponding effective diffusion time, and to

$$3f(1-f)\left(\langle \tilde{D}_{I,e}\rangle - \langle \tilde{D}_{I,a}\rangle\right)^2 \tag{32}$$

(red line).

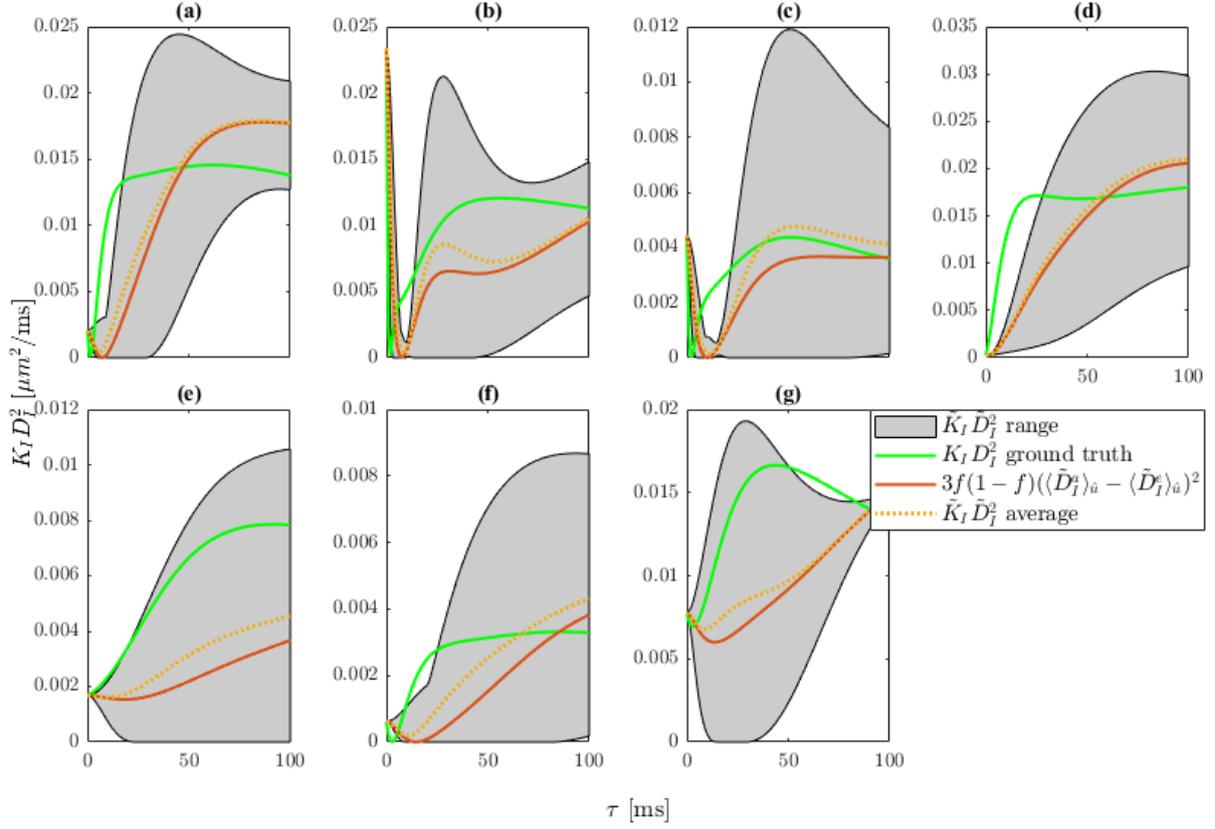

**Figure 4** For each of the 7 white matter ROIs a-g from [23], the orientational variability of isotropic kurtosis measurements as function of waveform (half) duration $\tau$ is illustrated. The gray shaded area in each plot defines the possible range of $\tilde{K}_I \tilde{D}_I^2$ values depending on the gradient frame orientation with respect to the spinal cord. The green curve plots the ground truth value $K_I D_I^2$ at the corresponding effective diffusion time, the full red curve plots (3 times) the variance of the directionally averaged compartment apparent isotropic diffusivities, and the dashed lines line the average $\tilde{K}_I \tilde{D}_I^2$ over directions.

It is evident that there is a large dependence on orientation of the apparent isotropic kurtosis in this system. Averaging $\tilde{D}_I^2 \tilde{K}_I$ over directions yields a measurement which for the most part is close to the variance of the average (over directions) apparent isotropic diffusivities, suggesting this to be the most accurate interpretation of the kurtosis in the system, assuming that intracompartmental kurtosis can be ignored. However, the average apparent isotropic diffusivities were seen in Figure 3 to be quite different from the actual isotropic diffusivities. Consistent with this, the orientationally averaged $\tilde{D}_I^2 \tilde{K}_I$ (dashed orange line) behaves very differently from the ground truth $D_I^2 K_I$ (green line).

Finally, Figure 5 illustrates the contribution to variance in apparent isotropic diffusivity originating from kurtosis in a single pore system, $\delta K_I^{(i)}$, as computed using Monte Carlo simulations. This contribution can become both positive and negative, depending on geometry, and in the case of the permeable barrier system, it becomes on the order of 1 for some characteristic length scales. Such a magnitude will clearly interfere with contributions from true ensemble variance in $D_I$ [47].

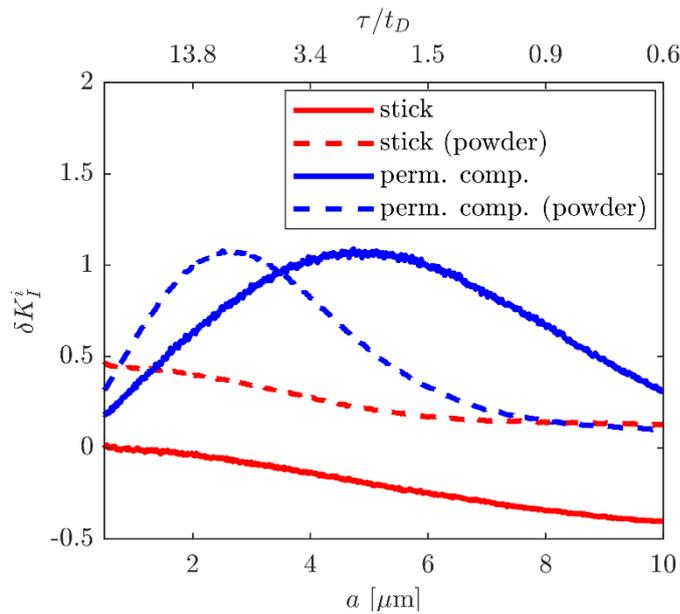

**Figure 5** *Here kurtosis $K_I = \delta K_I^{(i)}$ is plotted as function of barrier spacing for a single stick and the corresponding powder average (red), and for permeable barriers with $\kappa = 0.02 \mu m/ms$ (blue). The waveform was generated with $\tau = 55$ ms, but can be translated to other timings by using the upper x-axis.*

*Experiments*

To study the corruption of isotropic diffusivity measurements from a different angle, we now investigate the variability in apparent isotropic diffusivity in rat spinal cord as the q-MAS gradient frame is rotated along 12 orientations. Figure 6 shows raw signal decays from a few selected voxels along with fits to the DKI expression Eq. (19).

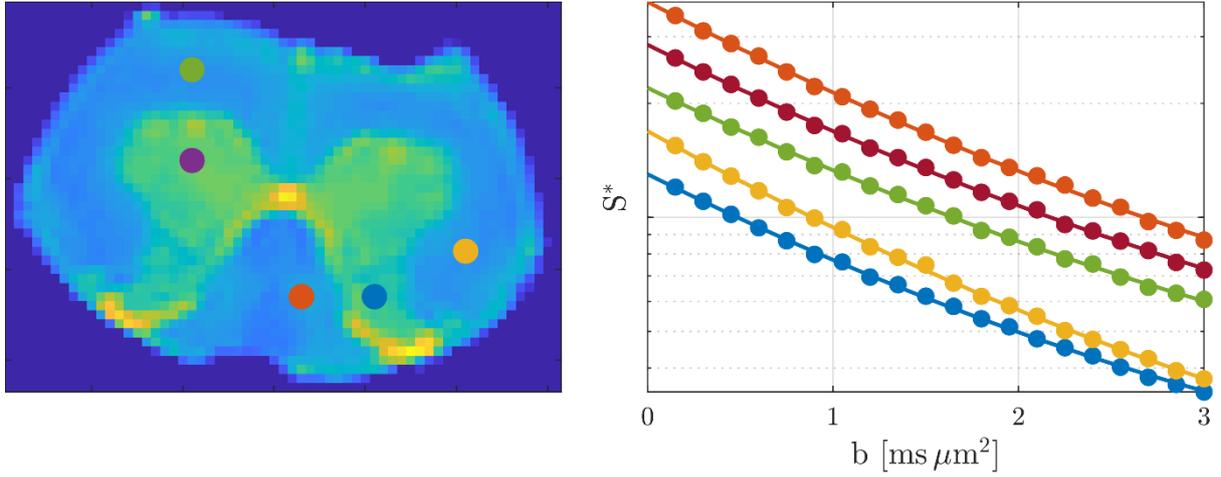

**Figure 6** Raw signal decay S* versus b from the voxels shown to the left. Each individual curve is scaled arbitrarily for visibility. Fits to Eq. (19) are shown with solid lines. Although the deviation from monoexponential decay appears small [47, 48], 99% of all voxels had $\tilde{K}_I > 0.2$.

In addition to a T2-weighted image $S_0$ and FA, Figure 7 shows maps of $\tilde{D}_I$ and $\tilde{K}_I$ standard deviation over orientations, normalized to their respective means over orientations. Note that 99% of all voxels had $\tilde{K}_I > 0.2$, so this normalization is unproblematic.

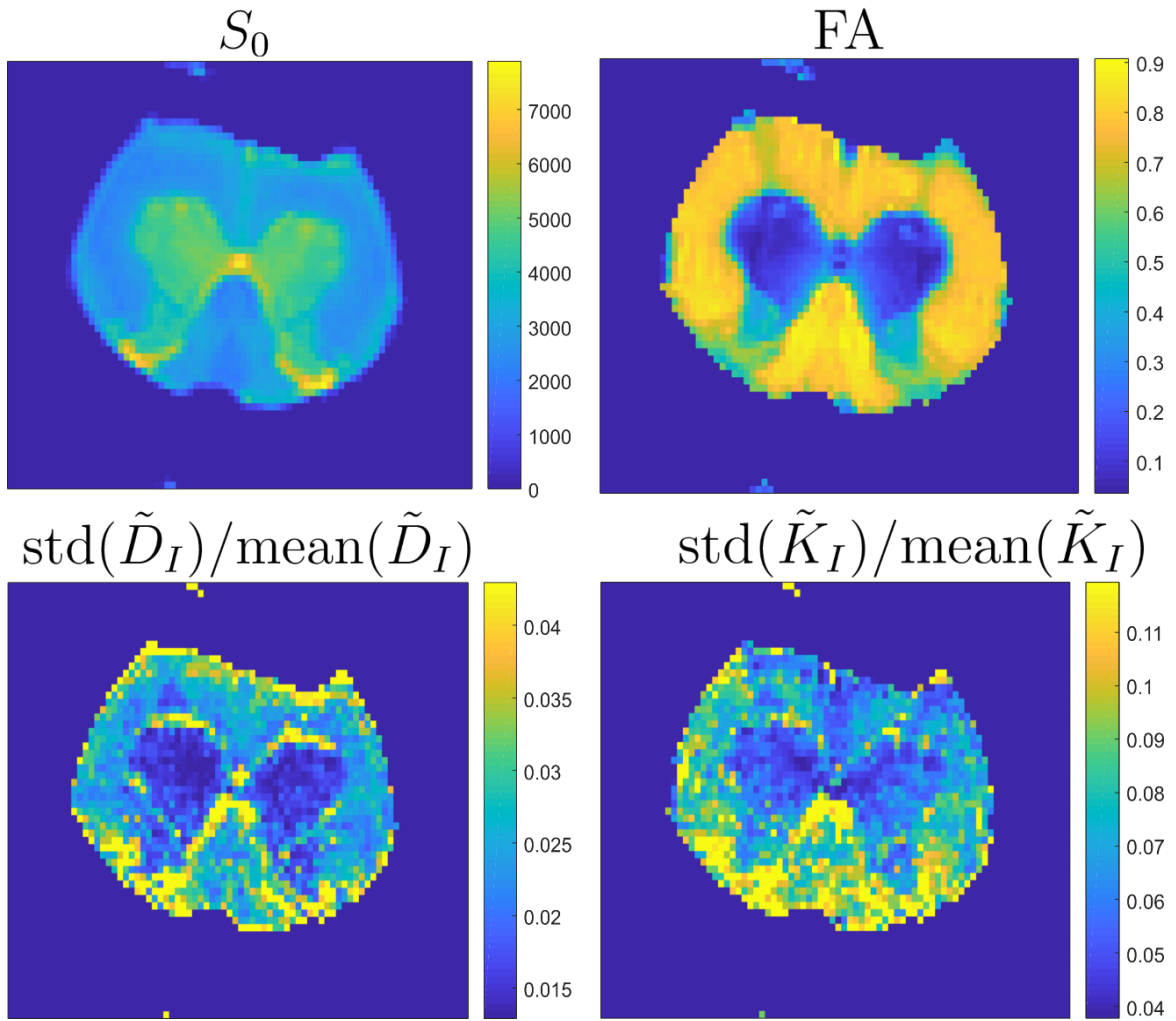

**Figure 7** *Parameter maps from the rat spinal cord experiment, $S_0$, fractional anisotropy FA, and standard deviation of $\tilde{D}_I$ (bottom left) and $\tilde{K}_I$ normalized by the mean (coefficient of variation).*

It is clear that the variability is quite small for the isotropic diffusivity, on the level of a few percent, whereas it is substantially higher for the kurtosis. The variability for both tends to be higher in the white matter, than in the gray matter, as could be expected. To isolate the effects of measurement uncertainty, we compare in Figure 8 and Figure 9 the variability over gradient frame orientations to the average variability over repetitions in white matter, identified here by $\text{FA} > 0.6$. In both cases, the variability over orientations is significantly larger than the variability over repetitions. However, for the isotropic diffusivity (Fig. 8a), the variability is quite low, with medians of 0.02 and 0.03 for repetitions and orientations, respectively. Nevertheless, it is clearly larger than in the phantom (Fig. 9b) where this variability is vanishing. On the other hand, in the case of the isotropic kurtosis, the median variabilities are 0.06 and 0.11 over repetitions and orientations, respectively.

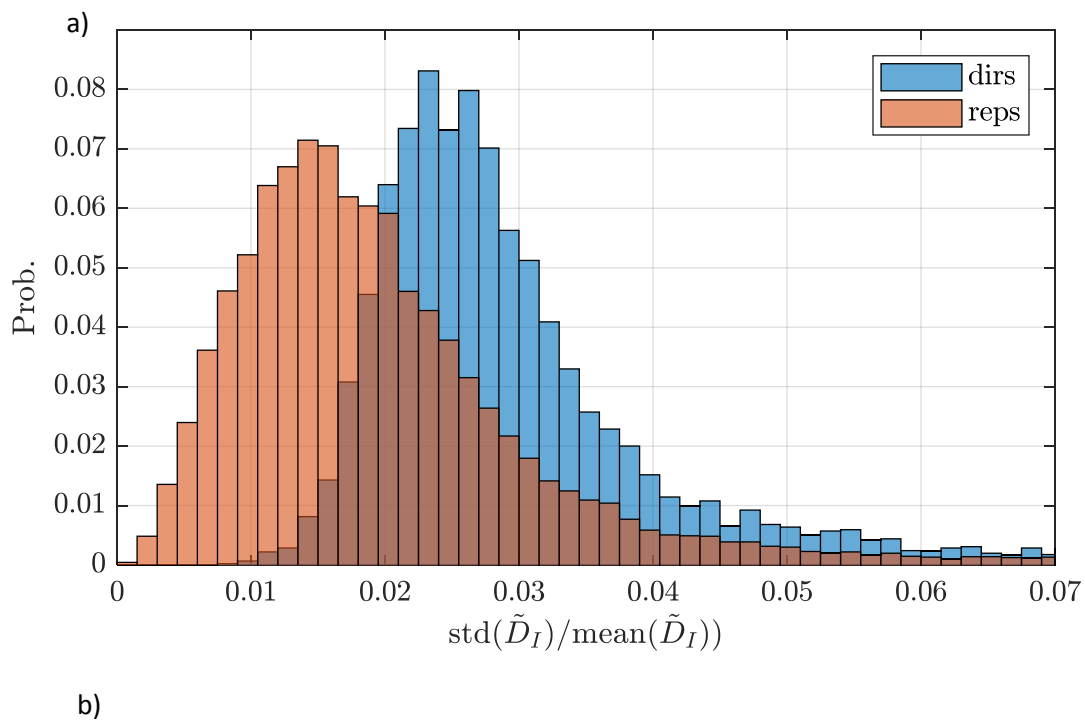

a)

b)

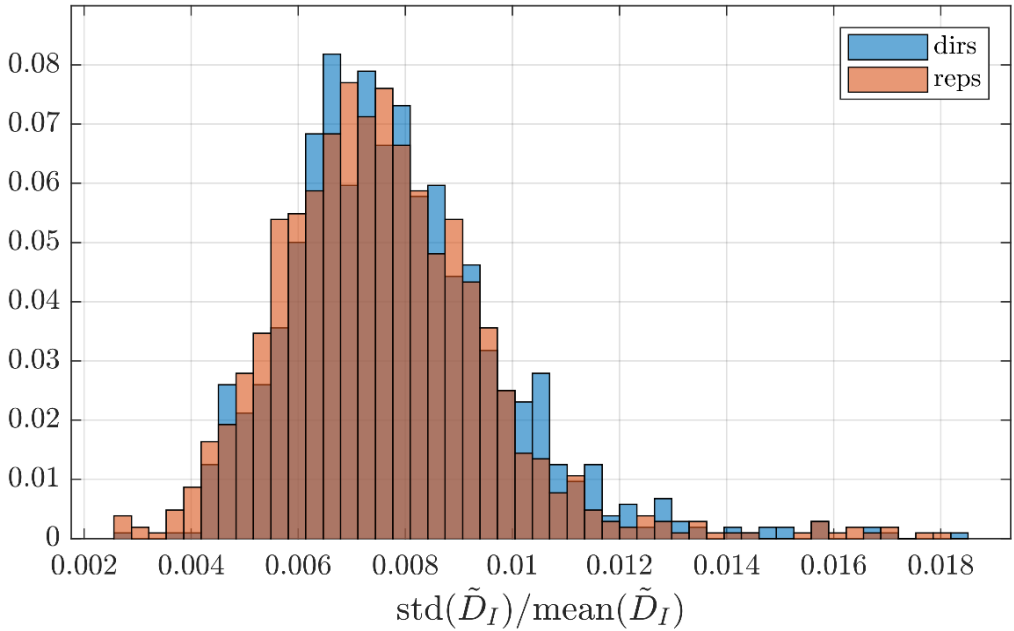

**Figure 8** *Histograms showing the variability of $\tilde{D}_I$ over the acquired gradient orientations (blue) and over repetitions (orange) in white matter (a), defined here as FA$>0.6$. The average and standard deviation over repetitions are 0.02 and 0.04, whereas both are 0.04 over orientations. The medians are 0.02 and 0.03 for repetitions and orientations respectively. For comparison, the same type of histogram is shown for the phantom in (b) showing no difference in variability due to orientations and repetitions.*

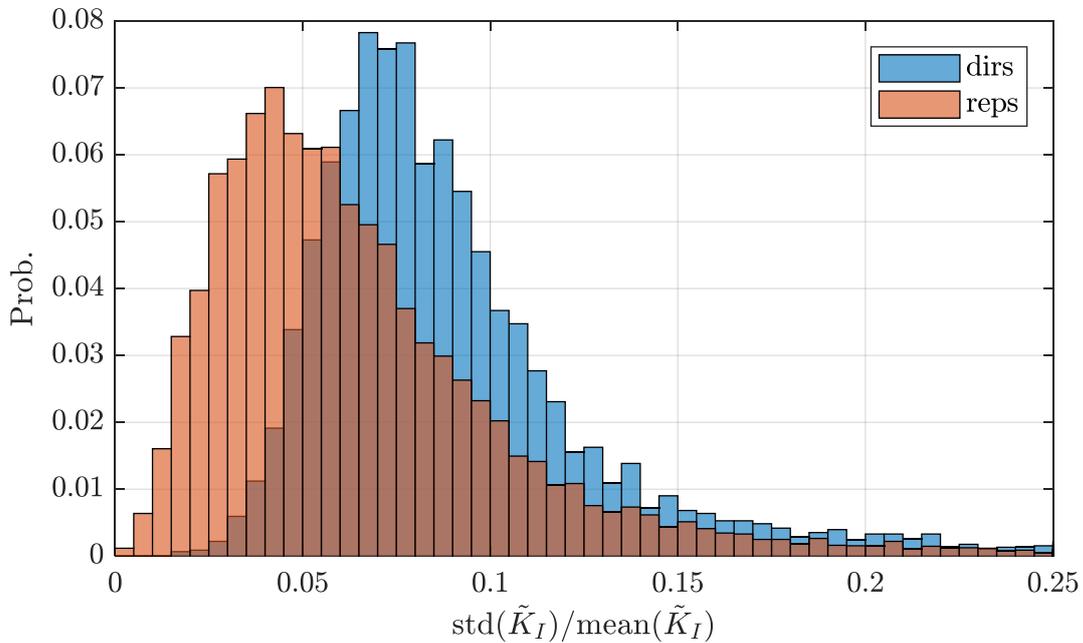

**Figure 9** *Histograms showing the variability of $\tilde{K}_I$ over the acquired gradient orientations in white matter, defined here as $FA > 0.6$. The average and standard deviation over repetitions are 0.08 and 0.1, whereas they are 0.11 and 0.15 over directions. The medians are 0.06 and 0.11 for repetitions and orientations respectively.*

## Discussion

In this paper, we addressed the general influence of nongaussian diffusion on the interpretation of multidimensional diffusion MRI. We focused on a particular realization of the so-called magic angle spinning of the q vector for isotropic diffusion weighting (q-MAS), and identified two potential confounds when the crucial assumption of multiple Gaussian components is violated. Both issues were shown to lead to additional contributions to the kurtosis otherwise identified exclusively with ensemble variance in isotropic diffusivity.

The first issue was the appearance of anisotropic time dependence in the diffusion tensor, which induced a dependence of apparent isotropic diffusivity on gradient frame orientation, as first consider theoretically by de Swiet and Mitra [36]. This causes an additional contribution to the q-MAS kurtosis due to pore orientation dispersion, which can jeopardize the interpretation of kurtosis as ensemble variance in isotropic diffusivity. The actual magnitude of the effect was found to depend on the geometry of the compartments, and their size compared to the diffusion length. Based on the square cylinder simulations, we found that for diameters below about 5 μm, the overall variability relative to the true bulk diffusivity was less than 10%. Such a geometry could apply, for example, to spins inside axons in the brain. However, in the spinal cord where axons are larger, the variability could become greater. For a uniform distribution of pores, the dispersion contribution to q-MAS kurtosis $\delta K_I^{(d)}$ remained below 0.07, or 12 to 28 % of previously quantified $\tilde{K}_I$ values, 0.25 – 0.6 [8, 46, 47], despite an apparently close to monoexponential decay [47, 48]. For a rectangular box, the simulations revealed a large class of geometries for which the range of variability in apparent isotropic diffusivity exceeded 0.1: in fact, for dimensions on the typical cellular scale, the pores had to be relatively isotropic to limit the range below 0.1, corresponding to 15 – 40% of reported $\tilde{K}_I$ values [8, 46, 47]. The corresponding standard deviation in a powder sample allowed for slightly larger anisotropy, but still limited the method to rather isotropic compartments. In the limit of completely isotropic compartments, the kurtosis of the standard Stejskal-Tanner sequence itself directly yields the variance in isotropic diffusivity over the ensemble (disregarding intracompartmental kurtosis, see below).

To examine the importance of these issues with plausible time-dependent compartmental diffusivities, we used results obtained in fixed pig spinal cord [23]. These results (Figs. 3-4) confirmed sufficient anisotropic diffusion time dependence to induce a small but noticeable orientation dependence of q-MAS $\tilde{D}_I$, in particular in intra-axonal space, as well as an overall overestimation of $D_I$. As a result of the anisotropic time dependence in each of these compartments, we observed a non-vanishing ensemble variance in isotropic diffusivity. Even though the orientational dependence of apparent isotropic diffusivity of the individual compartments in the system was small, the conflation with dispersion could potentially account for a large fraction of the isotropic variance. However, overall, the more serious issue for this type of geometry was the deviation between isotropic diffusivity and apparent isotropic diffusivity measured with q-MAS: this caused apparent isotropic kurtosis, even when averaged over orientations, to behave very differently from the ensemble variance of isotropic diffusivities. Furthermore, the apparent isotropic kurtosis showed a large orientational variability. These findings were explicitly confirmed with q-MAS acquisitions of different orientations in fixed rat spinal cord. There we found an orientational variability of $\tilde{K}_I$ typically about 11% of the mean, almost twice as large as the variability over repetitions. Note that this variability over gradient orientations is attenuated by intrinsic dispersion in the spinal cord — in the pig spinal cord, dispersion at comparable diffusion times was estimated to be approximately 20°[23].

The second issue was a non-vanishing intracompartmental kurtosis, which likewise adds to the ensemble heterogeneity in isotropic diffusivity ($\delta K_I^{(i)}$). Using Monte Carlo simulations, we analyzed its magnitude as a function of compartment size in the finite stick and regularly spaced permeable barriers in 1D. In these systems, the true ensemble variance of isotropic diffusivity is zero since there is just one type of compartment, possibly with different orientations. We found a rather large range of values from about -0.3 to more than 1 depending on the sample and its dimensions, which again could be compared to the previously quantified values of isotropic kurtosis in healthy brains and tumors falling in the range of 0.25 to 0.6 [8, 47, 56], although [48] reported only small deviation from mono exponential decay. This may potentially be a serious issue, not only for multidimensional diffusion weighting, but also for other models employing multiple Gaussian compartments [10, 57, 58].

In a given system, it may be possible to use the present results to design waveforms less sensitive to nongaussian effects, based on assumptions about pore geometry. Alternatively, one can apply multiple diffusion encoding with narrow pulsed field gradients, such as double diffusion encoding and triple diffusion encoding. With this approach, rigorous analysis shows that in the long mixing time regime, the $q^4$ (kurtosis) term directly yields the diffusion tensor variance, even for nongaussian pores [2, 15]. From this tensor, it is possible to extract isotropic diffusion variance directly [18, 59]. This approach may have other problems of a more practical nature, such as a prolonged echo time, but the regime of validity is arguably more transparent.

## Conclusion

Our results show that nongaussian diffusion effects confound the interpretation of multidimensional diffusion MR metrics, mainly due to the time-dependence of the diffusion tensor which is sampled in different and orientationally dependent ways during the gradient waveform application, as well as from intra-compartmental kurtosis. Consequently, caution should be exerted when interpreting metrics from multidimensional diffusion in nontrivial samples, as the interpretation of the derived parameters may be obscured by the abovementioned effects. Specifically, for magic angle spinning of the q-vector, where the diffusion kurtosis previously has been attributed solely to variance in isotropic diffusivity, we showed that additional contributions occur due to anisotropic time dependence of the diffusivity and due to kurtosis of individual pores. Experimental findings in fixed pig and rat spinal cords, using a 7.5 ms isotropic diffusion encoding waveform and 3000 mT/m maximum gradient strength, supported these conclusions.

## Acknowledgments

SNJ is supported by the Danish National Research Foundation (CFIN), and The Danish Ministry of Science, Innovation, and Education (MINDLab, Grant no. 0601–01354B). Funding from the European Research Council (ERC) under the European Union's Horizon 2020 research and innovation programme (Starting Grant, agreement No. 679058) supports NS, and funding from EPSRC grant number M507970 supports AI. We thank Dr. Daniel Nunes for tissue extraction.